%% file: Borkar_SiO-masers-in-GC_PoS.tex
\pdfoutput=1
\documentclass[11pt, useAMS, usenatbib]{PoS}
\usepackage{multirow}
\usepackage{graphicx}
\usepackage{multicol}
\usepackage{amssymb}
\usepackage{subcaption}
\usepackage[usenames,dvipsnames,svgnames,table]{xcolor}
\usepackage{lscape}
\usepackage{xtab, afterpage}


\title{Observations of the 86 GHz SiO maser sources in the Central Parsec of the Galactic Centre}
\ShortTitle{SiO Masers in the Galactic Centre}

\author{Borkar, A.$^{1}$\thanks{Speaker; E-mail: borkar@asu.cas.cz},
Eckart, A.$^{2,3}$, Straubmeier, C.$^{2}$, Sabha, N. B.$^{2}$, Sjouwerman, L. O.$^{4}$,
Karas, V.$^1$, Kunneriath, D.$^{5,1}$, Moser, L.$^{4,2}$, Britzen, S.$^{3}$, Valencia-S, M.$^{2}$,
Donea, A.$^{7}$, Zensus, A.$^{3,2}$
\vspace{0.4cm}\\
$^{1}$ Astronomical Institute of the Academy of Sciences, Prague, Bo\v{c}n\'{i} II 1401/1a, CZ-14131 Praha 4, Czech Republic\\
$^{2}$ I. Physikalisches Institut, Universit\"at zu K\"oln, Z\"ulpicher Strasse 77, D-50937, K\"oln, Germany\\
$^{3}$ Max-Planck-Institute f\"ur Radioastronomie, Auf Dem H\"ugel 69, D-53121, Bonn, Germany\\
$^{4}$ National Radio Astronomy Observatory, PO Box 0, Socorro, NM 87801, USA\\
$^{5}$ National Radio Astronomy Observatory, 520 Edgemont Road, Charlottesville, VA 22903, USA\\
$^{6}$ Argelander-Institut f\"ur Astronomie, University of Bonn, Auf dem H\"ugel 71, D-53121 Bonn, Germany\\
$^{7}$ Monash Centre for Astrophysics, Monash University, Clayton, Victoria 3800, Australia}

\abstract{We present results of 3 mm observations of SiO maser sources in the Galactic Centre (GC) from observations with the Australia Telescope Compact Array between $2010-2014$, along the transitions of the SiO molecule at $v = 1, J = 2-1$ at 86.243 GHz and $v = 2, J = 2-1$ at 85.640 GHz. We also present the results of the 3 mm observations with Atacama Large Millimeter/Submillimeter Array (ALMA). We detected 5 maser sources from the ATCA data, IRS 7, IRS 9, IRS 10EE, IRS 12N, and IRS 28; and 20 sources from the ALMA data including 4 new sources. These sources are predominantly late-type giants or emission line stars with strong circumstellar maser emission. We analyse these sources and calculate their proper motions. We also study the variability of the maser emission. IRS 7, IRS 12N and IRS 28 exhibit long period variability of the order of $1 - 2$ years, while other sources show steady increase or decrease in flux density and irregular variability over observation timescales. This behaviour is consistent with the previous observations.}

\FullConference{Multifrequency Behaviour of High Energy Cosmic Sources - XIII - MULTIF2019\\
		3-8 June 2019\\
		Palermo, Italy}

\begin{document}

\section{Introduction}

The innermost parsec of the Milky Way is a complex environment (hereafter referred to as the central parsec), with a supermassive black hole (SMBH) of mass of $4 \times 10^6 M_\odot$ at the centre. The SMBH is associated with the compact radio source, Sagittarius A* (Sgr A*) located at a distance of 8 kpc (see, e.g. \cite{Morris96, Melia01, Reid09, Genzel10, Eckart17} for detailed review of the topic).
The central parsec consists of an extremely dense collection of stars as well as `the mini-spiral' of filaments of ionized gas, and it is surrounded by a circumnuclear disk (CND) of neutral atomic \& molecular gas at $\simeq 1.5$ pc that extends upto $\sim 5 - 7$ pc. Observations of the central stellar cluster have discovered a large population of mainly late-type red giant stars, including asymptotic giant branch (AGB) stars (\textit{see e.g.} \cite{Ott99, Viehmann05, Paumard06, Tanner06, Peeples07}). Spectroscopic observations have found hot, early-type stars with H, He, and N emission lines showing characteristics of post main-sequence Ofpe/WN9, luminous blue variables (LBVs) and Wolf-Rayet stars \cite{Clenet01, Eckart04b, Moultaka05, Viehmann05, Viehmann06, Paumard06}.\let\thefootnote\relax\footnotetext{$^{\dagger}$ based on ALMA observations under the project 2013.1.00834.S executed on 10 April 2015 (PI: J. Darling).}

Recent observations have also discovered several massive young stellar objects (YSOs) ($\sim$$10 - 120~ M_{\odot}$, age $\lesssim 10 - 100$ Myr). These include more than 100 main-sequence OB stars, luminous OB giants and supergiants, post main-sequence Wolf-Rayet stars, and Herbig Ae/Be stars in the central parsec, as well as the S-star cluster of $>15$ main-sequence B stars within $1''$ ($0.04$ pc), very close to the SMBH \cite{Allen90, Krabbe91, Blum95a, Blum95b, Krabbe95, Tamblyn96, Najarro97, Ghez03, Paumard06, Lu09, YZ15, Eckart03, Eckart04b, Eckart05, Eckart13, Muzic08}. The presence of such young stars so close to the black hole is perplexing, and has been a hotly debated issue, sparking several dynamical models. These models are mainly categorised into two groups: (1) in-fall, or dynamical migration of the stellar cluster from outer zone (\cite{Kim-Morris03, Portegies-Zwart03, Wardle08} and references therein) and (2) \emph{in-situ} star formation (\cite{Mapelli12, Jalali14} and references therein). Analysis of the population of young stars in the central parsec suggests an age of $\sim$ $4 - 8$ Myr. The presence of two similar young (age $~2 - 7$ Myr), massive ($\sim\ 10^4 M_{\odot}$) star clusters with similar stellar population, the Arches and the Quintuplet, within 50 pc of the Galactic Centre (GC) alludes to the possibility of a global event a few million years ago that could have led to the star formation in the GC \cite{Figer02, Paumard06, Stolte15}. To understand the origins of the stars in the central parsec, observations of their spatial distribution in different frequency bands are required.

The central stellar cluster contains stars which are bright infrared (IR) sources and strong sources from the SiO maser emission. The SiO maser emission arises from the rotational transitions within the vibrational states of the SiO molecule, and is mainly associated with late-type red giant stars, especially AGB stars and planetary nebulae \cite{Nyman98, Desmurs00, Diamond03, Cotton04, Gray09, Perrin15}. SiO maser emission has been observed in massive young stellar objects, but it is very rare \cite{Elitzur92, Reid07_other, Matthews10, Goddi11}.
The radiative pumping from the IR radiation and collisional pumping are thought to be the mechanisms responsible for the SiO maser emission from the stellar atmosphere \cite{Gray09}. Very long baseline interferometry (VLBI) observations have shown that SiO masers originate in the circumstellar envelope of the star, at a distance of $\sim$$10^{14}$ cm, closer than the OH \& H$_2$O masers, which are found further out in the stellar atmosphere \cite{Diamond03}. The SiO emission arises from cells close to the star, which describes the stellar position within 1 mas and can be treated as point sources \cite{Elitzur92, Sjou98a}. Stellar objects with SiO maser emission act as reliable tracers for Galactic dynamics as they are not sensitive to non-gravitational perturbations \cite{Sjou02, Li10}.

Since these sources are strongly visible in both the radio line emission and the IR, their radio positions have been used to create a reference frame to perform accurate astrometry for the IR images which was used to locate the precise position of Sgr A* in IR \cite{Menten97, Reid03, Reid07}. The central parsec has been observed extensively for SiO sources at 43 GHz line at high resolution with the Very Large Array (VLA), the Very Long Baseline Array (VLBA), and the Australia Telescope Compact Array (ATCA) \cite{Menten97, Izumiura98, Miyazaki01, Deguchi00a, Deguchi00b, Deguchi02, Sjou02, Sjou04a, Imai02, Reid03, Reid07, Oyama08, Li10, YZ15}. Comparatively, the 86 GHz SiO line sources are still largely unexplored. \cite{Lindqvist91} and \cite{Messineo02} observed the inner Galaxy using single dish telescopes, i.e. Nobeyama 45 m telescope and IRAM 30 m telescope respectively, but their observations were limited by angular resolution and sensitivity. Interferometric observations at 86 GHz were performed by \cite{Li10} using ATCA in 2006 and 2008. Their observations were limited by the velocity coverage ($\pm 30 - 60$ km s$^{-1}$), where they detected two sources: IRS 10EE and IRS 15NE.\ Further observations by Li et al.\ also detected IRS 7, IRS 10EE, IRS 12N, IRS 15NE, IRS 17, IRS 28, SiO 15 and SiO 20 (J. Li, L. O. Sjouwerman \& C. Straubmeier, private communication). A census of the stellar sources in the central parsec at different wavelengths is necessary to understand the physical processes that govern the stellar distribution in the GC.

In this paper, we report the results of our observations of the GC environment at 3 mm, taken between 2010 to 2014 using the ATCA, and publicly available data from ALMA observation in 2015.\@ In Section 2, we describe the observations and data reduction processes used to obtain the positions of the detected maser sources. The results of the observations are discussed in Section 3, where we report proper motions of the stars. The analysis of the detected maser sources and their observed variability is discussed in Section 4, followed by a summary in Section 5.

\vspace{-0.2cm}
\section[]{Observations and Data Analysis}\label{Obs}

\noindent\textbf{ATCA Data:}

The observations of the Galactic Centre were made at 3mm band with the ATCA between 2010 and 2014. ATCA is an array of six 22-m telescopes located at the Paul Wild Observatory in Narrabri, NSW, Australia. Of the six antennas, five have 3 mm receivers. The location of ATCA allows us to observe the GC for more than 8 hours a day, as the GC passes almost overhead at the latitude of ATCA. The Compact Array Broadband Backend (CABB) was upgraded in 2007. The upgrades allow us to make observations with two wide 2048 MHz intermediate frequency (IF) bands at the spectral resolution of 1 MHz, which corresponds to the velocity resolution of $3.5$ km s$^{-1}$ and the velocity coverage of $\pm 3500~ \textup{km~s}^{-1}$. This wide band makes it possible to detect several high velocity maser sources. We performed the observations in a spectral line mode wherein we observed at two different frequency bands centred at 86.243 and 85.640 GHz. These frequencies correspond to two transition lines of the SiO molecule ($J = 2-1$, $v = 1$ and $J = 2-1$, $v = 2$). Observations were performed for approximately $10 - 12$ hours each day. The bandpass calibration with $\mathrm{PKS}$ $1253-055$ and flux calibration with Uranus were performed for 30 minutes each at the beginning and at the end of the observations, respectively. We observed the GC with three sets of 25 min on-source observations sandwiched between observations of gain calibrators (see Table \ref{Table1}). Sgr A* is a strong radio source thus it can be used for self calibration. These observations were carried out in part to study the variability of Sgr A* (see \cite{Borkar16}). For the 3 mm observations, the maximum available baseline is 214 m for the H214 configuration which was the predominantly used configuration. This gives primary beam of $30''$ and  synthesized beam of $1.99'' \times 2.28''$. The details of the observations are summarised in Table 1.

\begin{table}
\vspace{-0.75cm}
  \centering

\resizebox{\textwidth}{!}{%

  \begin{tabular}{l c c c c c c}
    \hline\hline
    \vspace{-0.2cm}
   & & & & & & \\
    & & & \multicolumn{2}{c}{Start Time} & \multicolumn{2}{c}{End Time}\\
   Date & Array & Calibrators & UT & JD & UT & JD\\[0.1cm]
    \hline
    \vspace{-0.2cm}
   & & & & & & \\
        & & & ATCA Observations &\\[0.1cm]
    2010 May 13 & H214 & 1741-312 & 11:04:45 & JD2455329.96042 & 21:50:25 & JD2455330.41001\\ 
    2010 May 14 & H214 & 1622-297 & 10:45:07 & JD2455330.948 & 22:07:40 & JD2455331.42199\\ 
    2010 May 15 & H214 & 1741-312 & 10:10:30 & JD2455331.92406 & 22:30:10 & JD2455332.42406\\ 
    2010 May 16 & H214 & 1622-297 & 10:08:47 & JD2455332.92277 & 21:35:60 & JD2455333.41961\\ 
    \vspace{0.1cm}
    \vspace{-0.3cm}
   & & & & & & \\
    2011 May 23 & H214 & 1741-312 & 09:57:43 & JD2455704.91516 & 21:05:13 & JD2455705.37862\\ 
    2011 May 24 & H214 & 1741-312 & 09:56:05 & JD2455705.91395 & 21:24:17 & JD2455706.39186\\ 
    2011 May 25 & H214 & 1741-312 & 10:01:13 & JD2455706.91751 & 21:22:30 & JD2455707.39062\\ 
    2011 May 26 & H214 & 1741-312 & 10:03:01 & JD2455707.91876 & 21:17:33 & JD2455708.38719\\ 
    \vspace{0.1cm}
    \vspace{-0.3cm}
   & & & & & & \\
    2012 May 15 & H214 & 1714-336 & 08:23:47 & JD2456062.84985 & 21:51:35 & J2456063.41082D\\ 
    2012 May 16 & H214 & 1714-336 & 10:04:30 & JD2456063.91979 & 21:45:22 & J2456064.4065D\\ 
    2012 May 17 & H214 & 1714-336 & 09:49:17 & JD2456064.90922 & 21:52:18 & JD2456065.41132\\ 
    2012 May 18 & H214 & 1714-336 & 11:02:51 & JD2456065.96031 & 21:56:46 & JD2456066.41442\\ 
   \vspace{0.1cm}
    \vspace{-0.3cm}
   & & & & & & \\
    2013 June 26 & EW352 & 1741-312 & 08:18:21 & JD2456469.84608 & 20:44:37 & JD2456470.36432\\ 
    2013 June 27 & EW 352 & 1741-312 & --- & --- & --- & ---\\ 
    2013 August 31 & 1.5A & 1741-312 & 03:36:08 & JD2456535.65009 & 14:57:04 & JD2456536.12296 \\ 
    2013 September 1 & 1.5A & 1741-312 & --- & --- & --- & ---\\ 
    2013 September 14 & H214 & 1741-312 & 03:08:12 & JD2456549.63069 & 13:27:588:12 & JD2456549.63069\\ 
   
    2013 September 16 & H214 & 1741-312 & --- & --- & --- & --- \\ 
    \vspace{0.1cm}
    \vspace{-0.3cm}
   & & & & & & \\
    2014 April 1 & H168 & 1741-312 & 14:54:24 & JD2456749.12111 & 23:38:55 & JD2456749.48536\\ 
    2014 April 2 & H168 & 1741-312 & 13:37:58 & JD2456750.06803 & 00:22:18 & JD2456750.51549 \\ 
    2014 June 7 & EW352 & 1714-336 & 08:00:50 & JD2456815.83391 & 19:00:02 & JD2456816.29169\\ 
    2014 September 26 & H214 & 1714-336 & 02:05:40 & JD2456926.58727 & 12:44:31 & JD2456927.03091\\ 
    2014 September 27 & H214 & 1714-336 & 02:10:13 & JD2456927.59043 & 12:40:51 & JD2456928.02837\\ 

    \vspace{-0.2cm}
   & & & & & & \\
    \hline
    \vspace{-0.1cm}
   & & & & & & \\
   & & & ALMA Observation &\\[0.1cm]
    2015 April 15 & C34-1/(2) & J1752-2956 & 06:16:33 & JD2457122.761493 & 07:14:33 & JD2457122.801771 \\
    \vspace{-0.15cm}
   & & & & & & \\
    \hline
  \end{tabular}}
    \caption{\small The Log of observations of Sgr A* taken using ATCA and ALMA. The dashes represent the days on which observations were not made. See Section 3 of \cite{Borkar16} for details.}\label{Table1}

\end{table}

The \texttt{MIRIAD} data reduction package was used to map and reduce the interferometer data \cite{Sault95}. The bandpass, gain and flux calibrations were performed following the standard procedure. We then perform a hybrid Hogbom/Clark/Steer \texttt{CLEAN} algorithm to produce a clean map from a dirty map. The task \texttt{UVLIN} was used to separate the continuum from line in the spectral data using a linear fitting to the line-free channels and the output was subtracted from the calibrated data. The continuum-subtracted data were then mapped as a spectral cube. The typical rms noise in the individual channels is $\sim 10$ mJy.\\

\noindent\textbf{ALMA Data:}

We also use dataset that was observed using ALMA from the project 2013.1.00834.S (PI: J. Darling) and executed on 10 April 2015. The ALMA array was in C34-1/(2) configuration with 35 antennas of the main 12-m antenna array, and baselines ranging from 13 to 356 metres. The spectral setup consisted of 4 spectral windows with 1920 channels, centred at 85.321, 86.251, 87.181 and 88.11 GHz, with a bandwidth of 937.5 MHz and spectral resolution of 0.5 MHz which corresponds to the velocity resolution of $\sim$1.7 kms$^{-1}$. The configuration gives primary beam of $55''$ and synthesized beam of $\sim 4.1 \times 2.5$ arcsec. The observations were made for 1 hour in total, starting with bandpass calibration with J$1717-3342$ and flux calibration with J$1733-1304$, and on source integration time of $\sim 38$ minutes with phase calibration between on source observation.

Basic data reduction and standard calibration was performed using the Common Astronomy Software Application (\texttt{CASA} v4.2, \cite{McMullin07}) with ALMA calibration pipeline. The continuum emission was subtracted from the calibrated data with a linear fitting to the line-free channels, and the spectral line map was produced using the \texttt{CLEAN} algorithm with natural weighting. The resulting image was corrected for primary beam to accurately represent the flux density of the detected sources. The obtained rms noise in the spectral channels is $\sim 2.5$ mJy.

Further data analysis was performed using \texttt{astropy} \cite{astropy} packages.\\

\noindent\textbf{Detecting the SiO maser sources:}

\noindent To detect SiO maser sources, first we calculated the rms noise $\sigma$ for each channel after which, we searched for all the instances where the flux density was more than $5 \sigma $. These resulted in several candidate maser sources. We put additional constraints to distinguish the genuine sources from artifacts: the candidate should be detected over several epochs and over several channels. We detect the H$^{13}$CN recombination line (rest frequency$=86.340$ GHz) in the ATCA data which is observed along the mini-spiral. The detections from the recombination line have broad spectral width, and show extended features. These are excluded from the search of SiO maser candidate sources. We also referred to the lists of previously known stellar sources in the central parsec observed at 43 GHz and in IR observations \cite{Viehmann05, Reid07, Li10} to verify the positions of detected candidates.

\begin{figure*}
  \centering
  \resizebox{0.99\textwidth}{!}{
    \subcaptionbox{\tiny{ATCA detections}}{\includegraphics[width=0.49\textwidth]{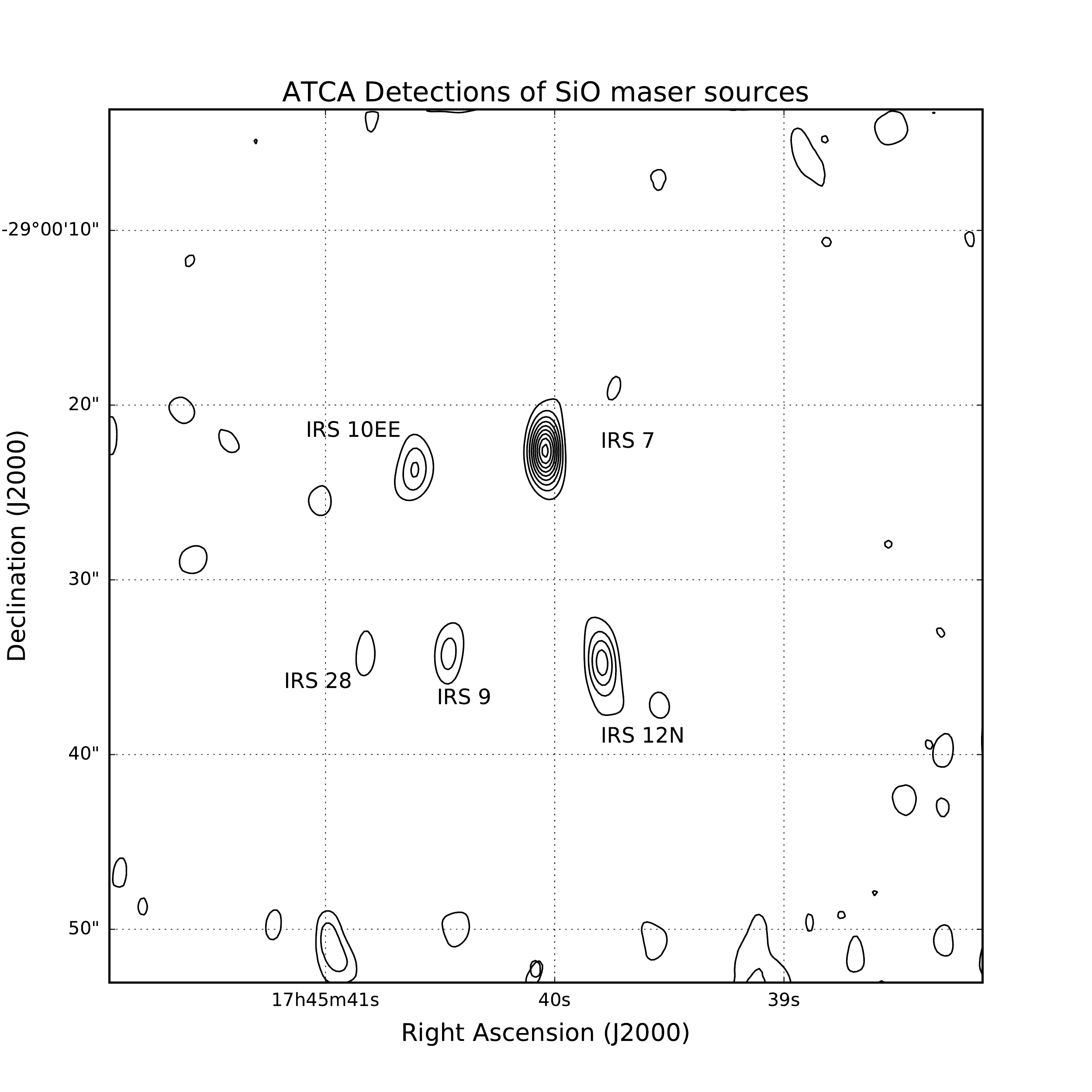}}
    \subcaptionbox{\tiny{ALMA detections}}{\includegraphics[width=0.49\textwidth]{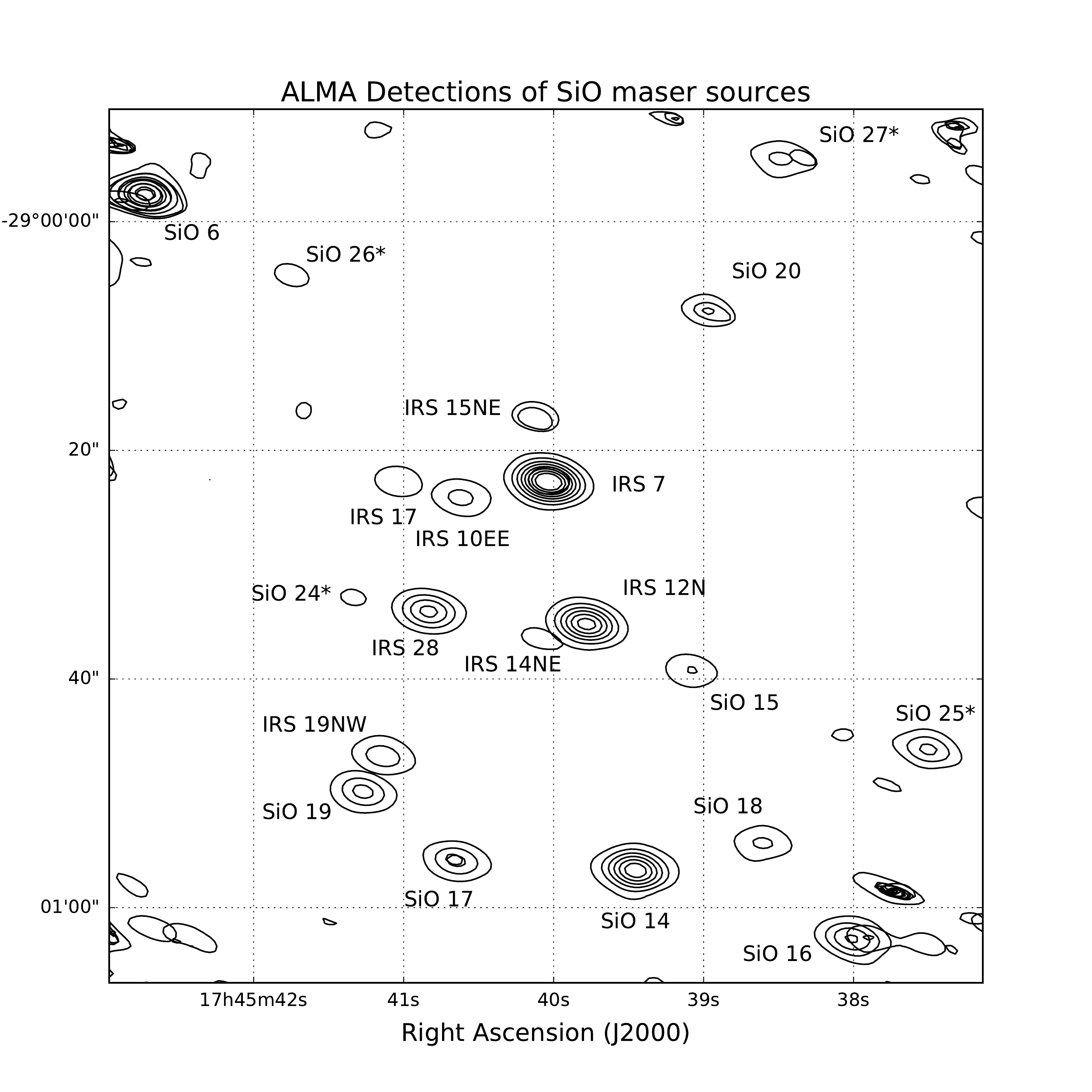}}}
  \caption{\small Positions of all the SiO maser sources detected with ATCA (\textit{left}) and ALMA (\textit{right}), respectively. Newly detected sources are marked with `*'. Contours are: \textit{left}: 0.05, 0.093, 0.137, 0.18, 0.225,0.268, 0.312, 0.356, 0.4 Jy; \textit{right}:  0.04, 0.09, 0.14, 0.19, 0.25, 0.29, 0.35, 0.4 Jy.}\label{SiO:all}
\end{figure*}

After this process, 5 sources were confirmed in the ATCA data and 20 sources in the ALMA data. The larger field-of-view and better sensitivity of ALMA facilitates the detection of weaker and greater number of sources further away from the centre. Of these, following are previously known sources: \textbf{ATCA data}: IRS 7, IRS 9, IRS 10EE, IRS 12N, and IRS 28; \textbf{ALMA data}: IRS 7, IRS 10EE, IRS 12N, IRS 14NE, IRS 15NE, IRS 17, IRS 19NW and IRS 28, and previously known 43 GHz SiO maser sources (\cite{Reid07, Li10}) SiO 6, SiO 14, SiO 15, SiO 16, SiO 17, SiO 18, SiO 19, and SiO 20. Four new sources were also detected in the ALMA data, which have been observed for the first time in radio emission lines. Following the nomenclature used by \cite{Reid07} and \cite{Li10}, we have named them SiO 24, SiO 25, SiO 26 and SiO 27 (See Fig.~\ref{SiO:all}).

\vspace{-0.2cm}

\begin{table}
  \vspace{-0.75cm}
  \centering
  \resizebox{0.8\textwidth}{!}{%

    \tiny
    
  \begin{tabular}{l r r r r}
    \hline
    \vspace{0.8mm}
    Source & Year & RA & DEC & Peak \\
           &      & ($''$) & ($''$) & Jy\\[0.8mm]
    \hline
    \vspace{-0.2cm}
           & & & &\\
           & & ATCA Detections & &\\
           & & & &\\[-0.1cm]
     IRS 7 & 2010.364383 & $0.0580 \pm 0.0365$ & $5.3279 \pm 0.0365$ & $0.685 \pm 0.032$\\
           & 2010.367123 & $-0.016 \pm 0.0366$ & $5.4459 \pm 0.0365$ & $0.788 \pm 0.02$\\
           & 2010.369863 & $0.0340 \pm 0.0365$ & $5.4019 \pm 0.0365$ & $0.654 \pm 0.01$\\
           & & & & \\[-0.1cm]                                          
           & 2011.394520 & $-0.096 \pm 0.0368$ & $5.3199 \pm 0.0366$ & $0.344 \pm 0.028$\\
           & 2011.397260 & $0.0920 \pm 0.0368$ & $5.3699 \pm 0.0367$ & $0.305 \pm 0.022$\\
           & 2011.400000 & $0.0860 \pm 0.0366$ & $5.4380 \pm 0.0365$ & $0.288 \pm 0.024$\\
           &      & & & \\[-0.1cm]                                           
           & 2012.372602 & $0.0320 \pm 0.0323$ & $5.5040 \pm 0.0354$ & $0.519 \pm 0.022$\\
           & 2012.375342 & $0.0180 \pm 0.0316$ & $5.3940 \pm 0.0326$ & $0.538 \pm 0.023$\\
           & 2012.378082 & $0.0599 \pm 0.0318$ & $5.5079 \pm 0.0331$ & $0.485 \pm 0.0224$\\
           & 2012.380821 & $0.0159 \pm 0.0326$ & $5.4720 \pm 0.0396$ & $0.467 \pm 0.0224$\\
           &      & & &\\[-0.1cm]                                            
           & 2014.252050 & $0.1639 \pm 0.0316$ & $5.3839 \pm 0.0316$ & $0.373 \pm 0.037$\\
           & 2014.432877 & $-0.114 \pm 0.0449$ & $5.2339 \pm 0.0749$ & $0.400 \pm 0.051$\\
           & 2014.736986 & $-0.080 \pm 0.0316$ & $5.3980 \pm 0.0316$ & $0.597 \pm 0.023$\\
           & 2014.739726 & $0.0840 \pm 0.0316$ & $5.4019 \pm 0.0316$ & $0.625 \pm 0.047$\\
           & & & &\\[-0.1cm]
    \vspace{-0.25cm}
           & & & &\\
     IRS 9 & 2010.364383 & $5.6852 \pm 0.0816$ & $ -6.3228 \pm 0.0816$ & $ 0.209 \pm 0.024$\\
           & 2010.367123 & $5.7602 \pm 0.0817$ & $ -6.1762 \pm 0.0816$ & $ 0.226 \pm 0.024$\\
           & 2010.369863 & $5.6314 \pm 0.0824$ & $ -6.1360 \pm 0.0819$ & $ 0.253 \pm 0.026$\\
           & & & &\\[-0.1cm]
           & 2011.394520 & $5.8888 \pm 0.0840$ & $ -5.9422 \pm 0.0824$ & $ 0.243 \pm 0.034$\\
           & 2011.397260 & $5.7668 \pm 0.0822$ & $ -6.0506 \pm 0.0819$ & $ 0.242 \pm 0.023$\\
           & 2011.400000 & $5.7024 \pm 0.0817$ & $ -6.1248 \pm 0.0817$ & $ 0.299 \pm 0.054$\\
           & & & &\\[-0.1cm]
           & 2012.372602 & $5.6090 \pm 0.0712$ & $ -6.2842 \pm 0.0718$ & $ 0.113 \pm 0.024$\\
           & 2012.375342 & $5.6382 \pm 0.0765$ & $ -6.1046 \pm 0.0883$ & $ 0.121 \pm 0.021$\\
           & 2012.378082 & $5.5764 \pm 0.0736$ & $ -6.0774 \pm 0.0790$ & $ 0.112 \pm 0.026$\\
           & 2012.380821 & $5.6736 \pm 0.0709$ & $ -6.2280 \pm 0.0732$ & $ 0.139 \pm 0.022$\\
           & & & &\\[-0.1cm]
           & 2014.736986 & $5.8190 \pm 0.1006$ & $ -6.1934 \pm 0.1001$ & $ 0.110 \pm 0.023$\\
           & 2014.739726 & $5.6662 \pm 0.1008$ & $ -5.7426 \pm 0.1007$ & $ 0.146 \pm 0.025$\\
           & & & &\\[-0.1cm]
    \vspace{-0.25cm}
           & & & &\\
    IRS 10EE & 2010.364383 & $7.6720 \pm 0.0817$ & $4.0779 \pm 0.0817$ & $0.378 \pm 0.0211$\\
           & 2010.367123 & $7.6776 \pm 0.0816$ & $4.2279 \pm 0.0816$ & $0.399 \pm 0.0220$\\
           & 2010.369863 & $7.5818 \pm 0.0816$ & $4.1920 \pm 0.0816$ & $0.381 \pm 0.0210$\\
           & & & &\\[-0.1cm]
           & 2011.394520 & $7.6126 \pm 0.0817$ & $4.1180 \pm 0.0817$ & $0.280 \pm 0.0240$\\
           & 2011.397260 & $7.6018 \pm 0.0818$ & $4.1420 \pm 0.0817$ & $0.244 \pm 0.0200$\\
           & 2011.400000 & $7.5806 \pm 0.0818$ & $4.1779 \pm 0.0817$ & $0.204 \pm 0.0180$\\
           & & & &\\[-0.1cm]
           & 2012.372602 & $7.5163 \pm 0.0720$ & $3.8980 \pm 0.0735$ & $0.217 \pm 0.0170$\\
           & 2012.375342 & $7.6503 \pm 0.0750$ & $4.2540 \pm 0.0960$ & $0.152 \pm 0.0130$\\
           & 2012.378082 & $7.5348 \pm 0.0723$ & $4.4220 \pm 0.0737$ & $0.144 \pm 0.0148$\\
           & 2012.380821 & $7.5770 \pm 0.0750$ & $3.9379 \pm 0.0825$ & $0.155 \pm 0.0144$\\
           & & & &\\[-0.1cm]
           & 2014.249315 & $7.4437 \pm 0.1415$ & $3.8639 \pm 0.1414$ & $0.120 \pm 0.0100$\\
           & & & &\\[-0.1cm]
    \vspace{-0.25cm}
           & & & &\\
    IRS12N &2010.364383 & $ -3.1940 \pm 0.0824$ & $-7.0684 \pm 0.0820$ & $0.100 \pm 0.015$\\
           &2010.367123 & $ -3.2439 \pm 0.0820$ & $-6.8922 \pm 0.0819$ & $0.143 \pm 0.015$\\
           &2010.369863 & $ -3.6560 \pm 0.0820$ & $-7.0163 \pm 0.0818$ & $0.160 \pm 0.017$\\
           & & & &\\[-0.1cm]
           &2012.372602 & $ -3.2560 \pm 0.0579$ & $-6.6343 \pm 0.0583$ & $0.260 \pm 0.017$\\
           &2012.375342 & $ -3.2959 \pm 0.0578$ & $-6.8578 \pm 0.0583$ & $0.223 \pm 0.016$\\
           &2012.378082 & $ -3.1279 \pm 0.0578$ & $-6.9894 \pm 0.0581$ & $0.200 \pm 0.016$\\
           &2012.380821 & $ -3.1100 \pm 0.0624$ & $-6.7728 \pm 0.0723$ & $0.230 \pm 0.017$\\
           & & & &\\[-0.1cm]
           &2014.736986 & $ -3.2500 \pm 0.0817$ & $-6.8633 \pm 0.0816$ & $0.480 \pm 0.033$\\
           &2014.739726 & $ -3.2960 \pm 0.0818$ & $-6.8612 \pm 0.0817$ & $0.490 \pm 0.030$\\
           & & & &\\[-0.1cm]
  \end{tabular}}
  \caption{\small Position offsets from Sgr A* and flux density of the detected SiO maser sources. `*' marks the newly discovered sources.}
  \label{SiO_data:ATCA}
\end{table}

\begin{table*}\label{SiO:ALMA}
  \tiny
  \centering
  \resizebox{0.8\textwidth}{!}{%
  \begin{tabular}{l r r r r}
    \hline
    Source & Year & RA & DEC & Peak \\
           &      & ($''$) & ($''$) & Jy\\[0.8mm]
           \hline
           \vspace{-0.25cm}
           & & & &\\
    IRS 28 & 2010.364383 & $9.9878  \pm 0.0828$ & $-5.7386 \pm 0.0820$ & $0.134 \pm 0.017$\\
           & 2010.367123 & $10.3094 \pm 0.0822$ & $-5.7244 \pm 0.0820$ & $0.110 \pm 0.014$\\
           & 2010.369863 & $10.4244 \pm 0.0817$ & $-5.8432 \pm 0.0817$ & $0.153 \pm 0.017$\\
           & & & &\\[-0.1cm]
           & 2012.372602 & $10.4798 \pm 0.0601$ & $-5.8726 \pm 0.0617$ & $0.070 \pm 0.015$\\
           & 2012.375342 & $10.3314 \pm 0.0633$ & $-5.4248 \pm 0.0763$ & $0.074 \pm 0.012$\\
           & 2012.378082 & $10.3444 \pm 0.0660$ & $-6.1092 \pm 0.0876$ & $0.073 \pm 0.012$\\
           & 2012.380821 & $10.0470 \pm 0.0644$ & $-5.8776 \pm 0.0766$ & $0.065 \pm 0.014$\\
           & & & &\\[-0.1cm]
           & 2014.432877 & $10.2916 \pm 0.0825$ & $-5.8406 \pm 0.0817$ & $0.103 \pm 0.018$\\
           & 2014.736986 & $10.2344 \pm 0.2097$ & $-5.8964 \pm 0.1725$ & $0.100 \pm 0.023$\\
           & 2014.739726 & $10.0848 \pm 0.0848$ & $-5.9746 \pm 0.0832$ & $0.084 \pm 0.015$\\
           & & & &\\[-0.1cm]
    \hline
           & & & &\\
           & & ALMA Detections & &\\
           & & & &\\
    IRS 7 & 2015.2726 & $0.0243 \pm 0.0063$ & $5.433 \pm 0.0039$ & $0.4945 \pm 0.003$\\
    IRS 10EE & 2015.2726 & $7.5930 \pm 0.0110$ & $4.0300 \pm 0.0067$ & $0.1102 \pm 0.003$\\
    IRS 12N & 2015.2726 & $-3.3214 \pm 0.0047$ & $-7.0085 \pm 0.0030$ & $0.3195 \pm 0.003$\\
    IRS 14NE & 2015.2726 & $0.6355 \pm 0.0288$ & $-8.2528 \pm 0.0154$ & $0.0300 \pm 0.003$\\
    IRS 15NE & 2015.2726 & $1.0402 \pm 0.0198$ & $11.0842 \pm 0.0127$ & $0.0818 \pm 0.003$\\
    IRS 17 & 2015.2726 & $13.0620 \pm 0.0128$ & $5.4520 \pm 0.0081$ & $0.0820 \pm 0.003$\\
    IRS 19NW & 2015.2726 & $14.3447 \pm 0.0156$ & $-18.5563 \pm 0.0093$ & $0.1293 \pm 0.003$\\
    IRS 28 & 2015.2726 & $10.4397 \pm 0.0072$ & $-5.8824 \pm 0.0043$ & $0.2149 \pm 0.003$\\
           & & & &\\[-0.1cm]
    SiO 6 & 2015.2726 & $35.0845 \pm 0.0192$ & $30.4963 \pm 0.0114$ & $0.3363 \pm 0.003$\\
    SiO 14 & 2015.2726 & $-7.5799 \pm 0.0097$ & $-28.5094 \pm 0.0061$ & $0.4364 \pm 0.003$\\
    SiO 15 & 2015.2726 & $-12.4102 \pm 0.0146$ & $-11.0621 \pm 0.0096$ & $0.0943 \pm 0.003$\\
    SiO 16 & 2015.2726 & $-26.4721 \pm 0.0182$ & $-34.4683 \pm 0.0112$ & $0.2036 \pm 0.003$\\
    SiO 17 & 2015.2726 & $7.9990 \pm 0.0171$ & $-27.7260 \pm 0.0098$ & $0.1602 \pm 0.003$\\
    SiO 18 & 2015.2726 & $-18.5807 \pm 0.0224$ & $-26.1446 \pm 0.0138$ & $0.1016 \pm 0.003$\\
    SiO 19 & 2015.2726 & $16.1361 \pm 0.0112$ & $-21.6761 \pm 0.0070$ & $0.1652 \pm 0.003$\\
    SiO 20 & 2015.2726 & $-13.9999 \pm 0.0161$ & $20.3122 \pm 0.0101$ & $0.0943 \pm 0.003$\\
           & & & &\\[-0.1cm]
    SiO 24* & 2015.2726 & $17.0794 \pm 0.022$ & $-9.2471 \pm 0.0142$ & $0.0496 \pm 0.003$\\
    SiO 25* & 2015.2726 & $-33.1316 \pm 0.0152$ & $-17.9372 \pm 0.0088$ & $0.1591 \pm 0.003$\\
    SiO 26* & 2015.2726 & $22.4607 \pm 0.0302$ & $23.5523 \pm 0.0192$ & $0.0571 \pm 0.003$\\
    SiO 27* & 2015.2726 & $-20.3771 \pm 0.009$ & $33.6992 \pm 0.0201$ & $0.1057 \pm 0.003$\\
           & & & &\\[-0.1cm]
    \hline
  \end{tabular}}
\end{table*}

\section[]{Results} \label{Results}

\noindent\textbf{Proper Motion Analysis:}

As discussed in section~\ref{Obs}, the longest baseline available for observation with ATCA is $\sim$ 214 m, and 356 m for the ALMA dataset, which translates to best possible angular resolution of $\sim 2''$. This is significantly lower resolution than the resolution available with VLBA at 43 GHz. To improve the accuracy in determination of the position of the source, we fit a two dimensional Gaussian to each detected source in each channel using \texttt{DPUSER} (\cite{Ott13}). We then obtain the position of the SiO maser for the observation day by weighted averaging the channel detections.

To calculate the proper motions of the masers, we assume that the stellar positions do not change within short time ($\sim$ days). We obtain the mean coordinates for a particular observation year by weighted averaging the positions obtained for each individual observation day. The 1$\sigma$ uncertainty in the position measurement is obtained by following \cite{Malkin13}. The weighted mean is given by $\bar{x} = {\Sigma_i w_i x_x}/{\Sigma_i w_i};~ (i=1,...,n)$, where $x_i$ are the position measurements, weights $w_i = 1/s_i^2$ and $s_i$ are measurement uncertainties in $x_i$. The estimated error in the weighted mean is $\sigma = \sqrt{\sigma_1^2 + \sigma_2^2}$ where
\[\sigma_1 = \frac{1}{\sqrt{\Sigma_i w_i}}; \textup{~~and~~} \sigma_2 = \sqrt{\frac{\Sigma_i w_i(x_i-\bar{x})^2}{(n-1)\Sigma_i w_i}} \]
 We then fit a weighted straight line through the resultant positions, along with the data from \cite{Reid07} and \cite{Li10} to obtain the proper motions. We increased the uncertainties in the position values for large reduced $\chi^2$ values. The detailed values of obtained positions of individual maser sources for all epochs can be found in table \ref{SiO_data:ATCA}. The scatter in the positions is consistent within the accuracy of angular resolution of 2$''$ of our ATCA and ALMA observations.

\begin{figure*}
  \centering
  \includegraphics{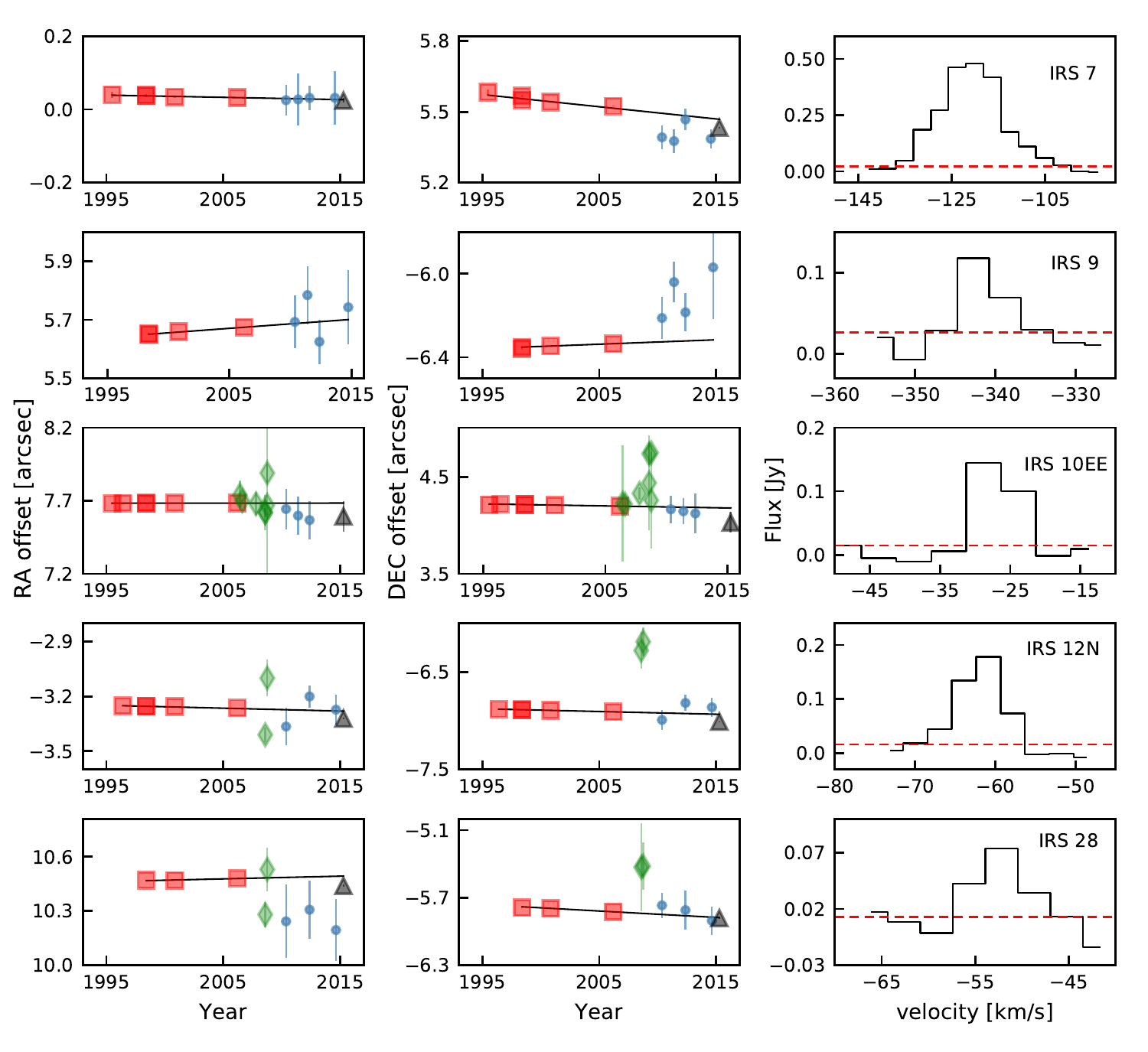}
  \caption{\small The RA and DEC proper motion and the observed ATCA spectra of IRS 7, IRS 9, IRS 10EE, IRS 12N, and IRS 28. In the columns on left and centre, the data from \cite{Reid07} and \cite{Li10} are represented by red squares and green diamonds, respectively. The blue circles represent data from our ATCA observations, and the results from ALMA data are represented by the black triangles. The red dashed line in right column shows the $1\sigma$ rms noise.}\label{SiO:prop-mot1}
\end{figure*}

\begin{figure*}
  \centering
  \includegraphics{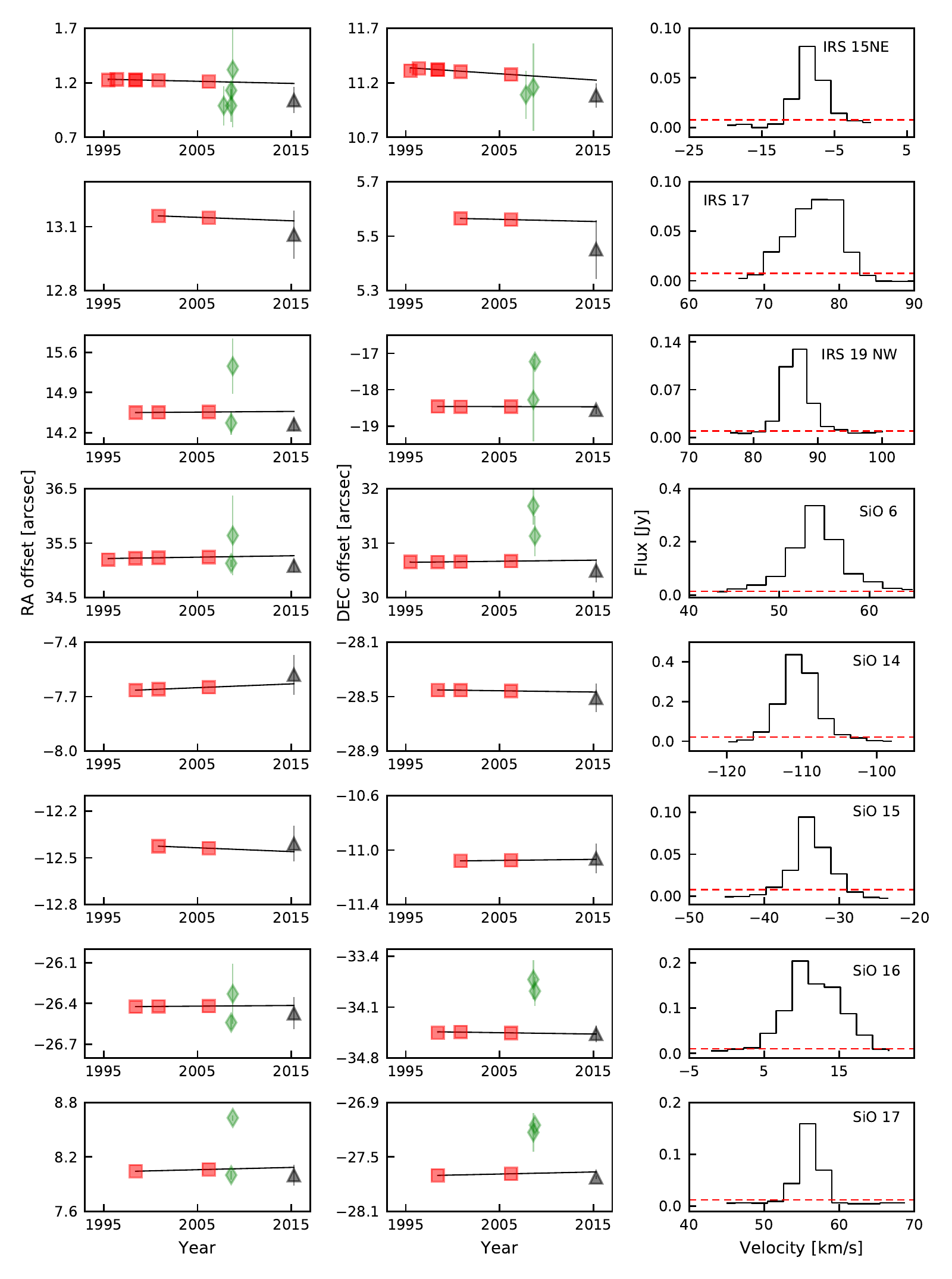}
  \caption{\small The RA and DEC proper motion and the observed ATCA spectra of IRS 7, IRS 9, IRS 10EE, IRS 12N, and IRS 28. In the columns on left and centre, the data from \cite{Reid07} and \cite{Li10} are represented by red squares and green diamonds, respectively. The blue circles represent data from our ATCA observations, and the results from ALMA data are represented by the black triangles. The red dashed line in right column shows the $1\sigma$ rms noise.}\label{SiO:prop-mot2}
\end{figure*}

Figure \ref{SiO:prop-mot1} shows the RA and DEC proper motions of the bright sources detected in the ATCA data with previously available data along with representative spectra from 17 December 2012. The red squares represent data from \cite{Reid07} and green diamonds represent data from \cite{Li10}, while the blue circles show points from our ATCA observations and the results from ALMA data are represented by the black triangles. The red dashed line in the images of spectra shows the $1 \sigma$ rms noise. Figure \ref{SiO:prop-mot2} shows the proper motions of the sources detected from the ALMA dataset. The proper motion values are converted to linear velocities, where the distance to the GC is assumed to be 8 kpc. Table~\ref{proper-motions} shows the values of the linear speed of the maser sources. These proper motions are consistent within the error margin with the previous studies.

\begin{figure*}
  \includegraphics{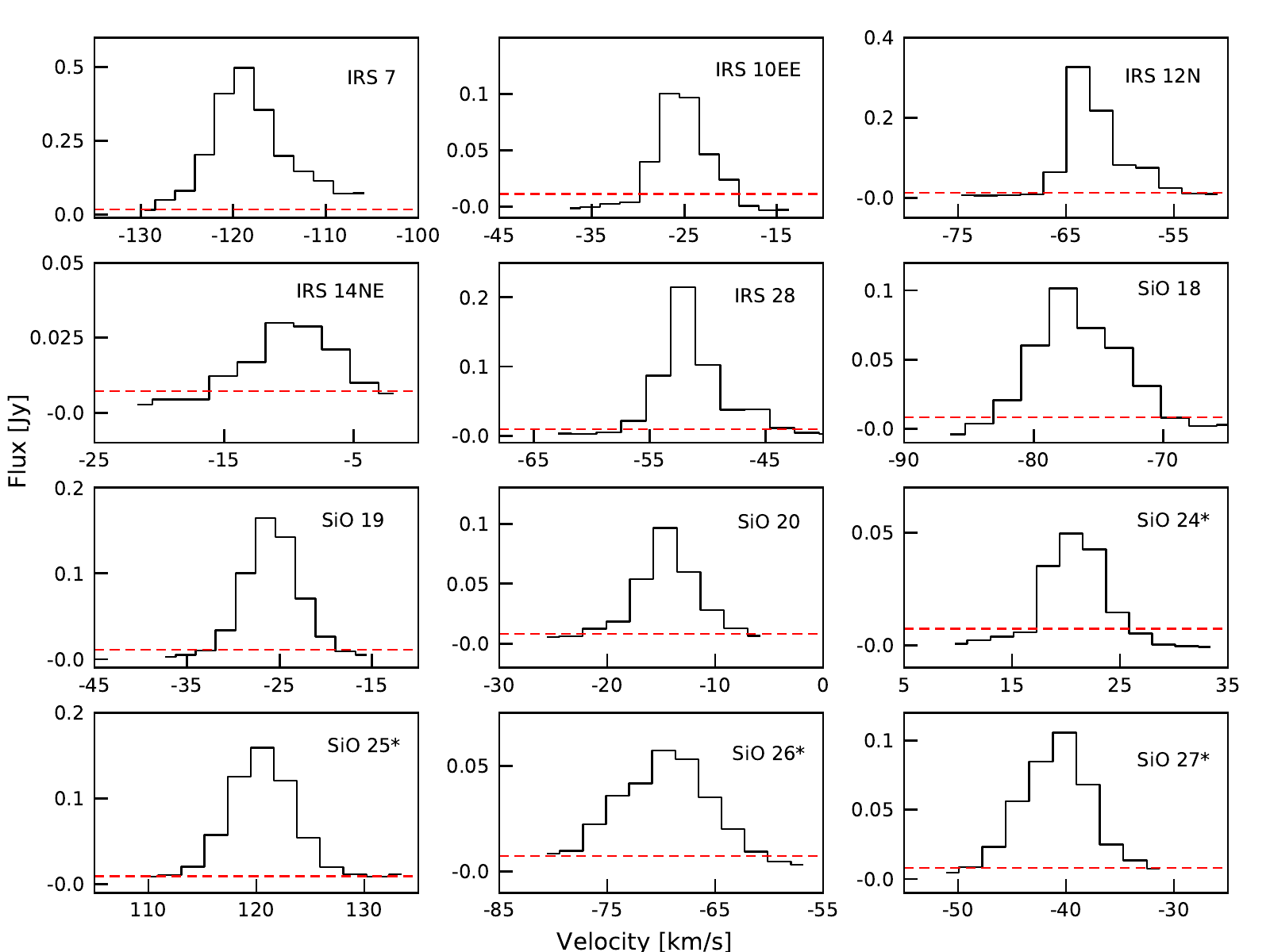}
  \caption{\small The spectra of the remaining sources detected with ALMA. The red dashed line shows the $1\sigma$ rms noise. The newly detected sources are marked with `*'}\label{SiO:ALMA_fl}
\end{figure*}
  
\begin{table}
  \vspace{-0.2cm}
  \centering
  \tiny
  \resizebox{0.95\textwidth}{!}{%
\begin{tabular}{l r r r r r}

\hline\\[0.1cm]

Source ID & $V_{LSR}$ & $\mu_{RA}$ & $\mu_{DEC}$ & $V_{RA}$ & $V_{DEC}$\\
  & $km/s$ & $mas/yr$ & $mas/yr$ & $km/s$ & $km/s$\\[0.1cm]
\hline\\[0.05cm]

  IRS7      & $-120 \pm 4$ & $-0.58 \pm 0.10$ & $-3.53 \pm 0.41$ & $-22.12 \pm 3.9$ & $-133.82 \pm 15.4$ \\
  IRS9 	    & $-340 \pm 4$ & $3.06 \pm 0.07$ & $2.11 \pm 0.24$ & $116.02 \pm 2.6$ & $80.10 \pm 9.3$ \\
  IRS10EE   & $-29 \pm 4$  & $0.03 \pm 0.04$ & $-2.09 \pm 0.06$ & $1.29 \pm 1.6$ & $-79.21 \pm 2.2$ \\
  IRS12N    & $-62 \pm 4$  & $-1.57 \pm 0.27$ & $-2.80 \pm 0.48$ & $-59.62 \pm 10.1$ & $-106.2 \pm 18.1$ \\
  IRS28     & $+18 \pm 4$  & $1.49 \pm 0.74$ & $-5.71 \pm 0.27$ & $56.67 \pm 28.1$ & $-216.4 \pm 10.4$ \\
  IRS15NE   & $+68 \pm 4$  & $-1.96 \pm 0.06$ & $-5.68 \pm 0.11$ & $-74.24 \pm 2.2$ & $-215.33 \pm 4.1$ \\
  IRS17     & $+77 \pm 4$  & $-1.62 \pm 0.14$ & $-0.79 \pm 0.27$ & $-61.55 \pm 5.4$ & $-29.83 \pm 10.2$ \\
  IRS19NW   & $-53 \pm 4$  & $1.18 \pm 0.22$ & $-0.43 \pm 0.28$ & $44.97 \pm 8.5$ & $-16.31 \pm 10.6$ \\
  SiO6 	    & $+43 \pm 4$  & $2.57 \pm 0.39$ & $1.97 \pm 0.35$ & $97.41 \pm 14.9$ & $74.88 \pm 13.3$ \\
  SiO14     & $-121 \pm 4$ & $2.08 \pm 0.30$ & $-0.94 \pm 1.95$ & $78.93 \pm 11.5$ & $-35.53 \pm 73.9$ \\
  SiO15     & $-44 \pm 4$  & $-2.46 \pm 0.27$ & $0.77 \pm 0.52$ & $-93.2 \pm 10.1$ & $29.29 \pm 19.7$ \\
  SiO16     & $+1 \pm 4$   & $0.49 \pm 0.07$ & $-1.84 \pm 1.08$ & $18.65 \pm 2.6$ & $-69.62 \pm 40.9$ \\
  SiO17     & $+47 \pm 4$  & $2.53 \pm 0.07$ & $2.25 \pm 0.66$ & $95.87 \pm 2.8$ & $85.55 \pm 25.1$ \\
  \hline
\end{tabular}}
\caption{\small SiO maser proper motions of the detected sources. The linear velocities are computed assuming 8 kpc distance to GC.}\label{proper-motions}

\end{table}

\vspace{-0.2cm}
\section[]{Discussion}

In this section we discuss the properties of the individual detected SiO maser sources as inferred from the analysis of their spectra. The central stellar cluster of the GC has been observed at various wavelengths over several decades. The multiwavelength observations, especially in near-infrared (NIR) and mid-infrared (mid-IR) wavelength, have been used for not only determining the precise positions of the IR sources (IRSs), but they have also been used to infer their spectral energy distributions (SEDs), and from these, their morphology has been determined. Here we give a brief overview of these studies.\\

\noindent\textbf{IRS 7:}

IRS 7 is a very strong radio SiO maser and IR source which has been used for adaptive optics guiding for IR observations. It is classified as M2 \cite{Perger08} or M1 {\cite{Paumard14}} type red super giant with bright thermal radio emission from its external envelope. It has a supergiant luminosity with its SiO maser features spread over more than $20$ km s$^{-1}$, which means that its envelope is much larger than typical Mira variable stars, with strongest features spanning about 10 mas. \cite{Paumard14} detect a long period variability of 2620-2850 days and a shorter period of $470 \pm 10$ days. The detected variation in the flux of IRS 7 in our dataset is consistent with the latter.\\ 

\noindent\textbf{IRS 9:}

IRS 9 is a late-type giant {\cite{Blum96}} shown to have characteristics of a Mira variable star {\cite{Reid07}}, of spectral type M3~ III {\cite{Figer03}}. It exhibits a very high three-dimensional velocity, exceeding the escape velocity at its projected distance. A number of explanations were proposed (see \cite{Reid07} for detailed analysis) to explain the observed high velocity. A binary system was postulated to explain its high velocity, but \cite{Reid07} show that the contribution to the velocity from the binary could not be more than $20$ km s$^{-1}$, and thus it could not explain the high velocity. Other explanations included existence of $0.8 \times 10^6$ M$_{\odot}$ dark mass in the form of stellar remnants, the distance to the GC exceeding 9 kpc, a non-zero $V_{LSR}$ of Sgr A*, or that IRS 9 is not bound to the GC. Their most-likely explanation is that it is a high-velocity star ejected from the central cluster. Calculations by \cite{Trippe08} suggest that although IRS 9 has high velocity, it is not excessive with respect to the global distribution.

IRS 9 is known to have a long period variability \cite{Blum96}. \cite{Peeples07} classify it to be large amplitude irregular or semi-regular variable, and \cite{Reid07} also consider it to be long period variable. In our ATCA data, it shows a variation in 86 GHz SiO emission over timescale of $\sim$ 2 years, which is consistent with the long period variability. IRS 9 was not detected in the ALMA data, possibly due to contamination from the surrounding ISM.\\

\noindent\textbf{IRS 10EE:}

From the IR observations, IRS 10EE has been shown to be a long-period variable star, of spectral type M1 III Mira variable, with variability ranging from several days to few months {\cite{Tamura96, Wood98, Ott99, Peeples07, Zhu08}}. It has also been observed at different maser lines, such as SiO, OH, and H$_2$O \cite{Lindqvist90, Lindqvist92a, Menten97, Sjou02, Reid03, Reid07, Oyama08, Peeples07}. The source also has a very compact SiO emission. It has been proposed that it could be a binary system \cite{Peeples07}, which has been supported by VLBA observations {\cite{Oyama08}}. \cite{Li10} detect variation on the timescale of months at 86 GHz. Since our observations are spaced every one year, this shorter variability cannot be detected in our data. We detect a steady decline in the peak flux density over the observation duration. \cite{Li10} detect peak flux density between 0.24$ - $0.45 Jy at 86 GHz, which suggests a possibility of $>5$ year long timescale variation.\\

\noindent\textbf{IRS 12N:}

IRS 12N has been observed in both IR and SiO line in radio, though it has a discrepancy between the IR and radio motions, possibly due to confusion caused by blending with other stars in the IR band. Similar to IRS 10EE, it has a very compact emission, and it has been classified as a cool red giant star {\cite{Blum96, Viehmann05}}, possibly a M0 III type Mira variable star {\cite{Figer03}}. It shows a steady increase in the flux density in the ATCA data, but a decrease in the ALMA data suggesting a possible longer variability period. Similar longer timescale variation has been observed at 43 GHz by \cite{Reid07}.\\

\noindent\textbf{IRS 14NE:}

IRS 14NE is an AGB star which has been observed in both IR and SiO line in radio and is often referred to as IRS 14N. It has been classified as M7 III type cool red giant \cite{Blum96} with possible long period variability (LPV) \cite{Blum03}. {\cite{Figer03}} classified it as M4 III late-type star. \cite{Ott99} and \cite{Peeples07} find it to have a \textit{K}$-$ band variability of $\sim 0.14$ magnitude and clear structure to the variability. Detected with a flux density of 30 mJy over only one spectral channel, IRS 14NE is the weakest SiO maser source detected in our ALMA dataset.\\

\noindent\textbf{IRS 15NE:}

The observations of IRS 15NE have lent mixed identification results. It has been identified as cool AGB star by \cite{Genzel96, Genzel00} while \cite{Krabbe95}, \cite{Paumard06}, and \cite{Martins07} classified it as WN9/Ofpe star. \cite{Blum03} found characteristic cool star features as well as strong He I and Br$\gamma$ emission in the spectrum, suggesting that the IR spectrum could be blended with the emission from a massive hot star. \cite{Najarro97} consider it to be a ``transition'' object, appearing as multiple sources. The 43 GHz VLBA observations by \cite{Oyama08} observed slightly resolved SiO emission and the small maser emission size to be compatible with the AGB star classification. ATCA observations by \cite{Li10} have shown that IRS 15NE is variable at both 43 and 86 GHz. It also exhibits high three-dimensional velocity.\\

\noindent\textbf{IRS 17:}

IRS 17 has been observed in both IR and radio line emission extensively over decades, and has been used to obtain accurate astrometry of the GC \cite{Reid03, Reid07, Schodel09, Gillessen09a, Plewa15}. \cite{Rieke78} and \cite{Lebofsky82} classified it as a possible red supergiant and \cite{Rieke78} found it to be bluer than IRS 7. It is classified as a late-type star by \cite{Feldmeier-Krause17}. \cite{Blum03} consider it to be a possible LPV star.\\

\noindent\textbf{IRS 19NW:}

There have been very few observations of IRS 19NW in both IR and radio. It has been observed at 43 GHz by \cite{Reid03} and \cite{Reid07} with VLA and VLBA, and by \cite{Li10} with ATCA, and in IR by \cite{Trippe08} and \cite{Plewa15}. IRS 19NW is one of the fainter sources in IR and is dominated by its bright neighbour IRS 19 \cite{Plewa15}.\\

\noindent\textbf{IRS 28:}

There have been several observations of IRS 28. \cite{Zhu08} classified it as M6~III type red giant. \cite{Tamura96} suggested that it is likely to be a variable while \cite{Blum96} classified it as LPV star. \cite{Peeples07} find it to ``not be clearly periodic'', and rule out a period of $\sim$200 days observed by \cite{Glass01}. It is visible every other year in our dataset and shows a variation in flux from $\sim 110 - 150$ mJy in 2010 to $\sim 70$ mJy in 2012 to $\sim 85 - 100$ mJy in 2014, suggesting that it may have a long period of a couple of years. It also shows comparatively large three-dimensional velocity.\\

\noindent\textbf{The `SiO-' sources:}

\cite{Reid07} and \cite{Li10} have detected several new SiO maser sources at 43 GHz apart from the well known IR sources (IRSs), with a combined total of 23 new sources. These sources are not detected in our ATCA images. Some of the sources are outside the field of view, and thus are not observed. For the sources that are within the field of view, the peak flux densities at 43 GHz range from 0.015 Jy to $\sim$ 0.1 Jy. Since the 43 GHz transition is usually (though not always) stronger than the 86 GHz transition \cite{Nyman93}, the corresponding peak flux density for these sources would be much lower at 86 GHz, where it would fall within $1-3 \sigma$ of the rms noise, where 1$\sigma ~=~ 0.015-0.030$ Jy. Thus most of these sources are not detected in our ATCA dataset. In the ALMA dataset, we identified 8 maser sources from the \cite{Reid07} and \cite{Li10} 43 GHz datasets: \textit{SiO-6, SiO 14, SiO 15, SiO 16, SiO 17, SiO 18, SiO 19}, and \textit{SiO 20}. Four other sources were also observed, which represent new detections. Following the nomenclature of \cite{Reid07} and \cite{Li10}, they were named as \textit{SiO 24, SiO 25, SiO 26}, and \textit{SiO 27}. Some of these sources have been previously observed at other wavelengths, though this is the first time they have been detected at mm wavelength. SiO 27 is a newly detected source which has not been observed at any other wavelengths. These newly detected sources are among the weaker sources detected in our ALMA data, with SiO 24, and SiO 26 having peak flux density $\sim 50$ mJy. Only IRS 14NE shows lower peak flux density.

\textbf{SiO 6} has been identified with V4928 Sgr and PSD J174542.72-285957.4 \cite{Peeples07} and is associated with OH, H$_2$O and SiO maser source OH 359.956-0.050 \cite{Levine95, YZ95, Sjouwerman96, Sjouwerman98b}. A long period variable IRS 24 has been suggested as its IR counterpart \cite{Levine95, Blum96}, although they are separated by large projected distance \cite{Peeples07}. Analysis by \cite{Sjouwerman96} suggests that it is an evolved intermediate mass AGB star. \cite{Glass01} and \cite{Peeples07} have observed large amplitude variability in the IR and is considered to be LPV candidate. \cite{Dong17} (\textit{ID 11}) suggest it to have a variability on a yearly time scale.

\textbf{SiO 16} is associated with V4911 Sgr, a LPV star \cite{Matsunaga09}. It has been observed to have a period of $\sim$ 528 days \cite{Glass01} while \cite{Peeples07} found it to have irregular or semi-regular variability without signs of periodicity. \cite{Dong17} (\textit{ID 451}) consider it to have a long period variability on yearly scale.

We identify \textbf{SiO 17} with OH 359.938-0.052 \cite{Sjouwerman98b} which is a known OH/IR star. \cite{Dong17} (\textit{ID 1933}) find it to be a non-variable star.

\textbf{SiO 20} can be identified with PSD J174538.98-290007.7 from \cite{Peeples07}. They observe a periodicity of $\sim 325$ days with large amplitude variations in both \textit{H-} and \textit{K-}bands, and consider it to be LPV star which is corroborated by \cite{Matsunaga09} and \cite{Dong17} (\textit{ID 1039}).

\textbf{SiO 25} is associated with V4910 Sgr (PSD J174537.24-290045.7 of \cite{Peeples07}) which is identified as a variable Star of Mira Cet type \cite{Matsunaga09}. Similar to SiO 16, a long period of 601 days was observed by \cite{Glass01} but \cite{Peeples07} found it to be irregular or semi-regular. A long timescale variability is detected by \cite{Dong17} (\textit{ID 212}).

Several `SiO-' sources have been discovered by \cite{Peeples07}, which are variable sources but do not show any periodicity. These are the following: \textbf{SiO 14} has been identified with PSD J174539.45-290056.6 \cite{Peeples07} which is SSTGC 523553 from \cite{Matsunaga09} and \textit{ID 522} of \cite{Dong17}. \cite{Peeples07} found it to be a blended star without any periodicity, but \cite{Blum96}, \cite{Blum03}, and \cite{Matsunaga09} consider it to be late-type LPV star, and \cite{Dong17} also find a long timescale variability. \textbf{SiO 15} is considered to be PSD J174539.09-290039.3 \cite{Peeples07}, \textit{ID 220} \cite{Dong17}. \cite{Blum96} consider it as a LPV candidate, and \cite{Dong17} find a yearly timescale variability. We identify \textbf{SiO 18} as PSD J174538.61-290054.3 \cite{Peeples07}. It is classified as a late-type \cite{Feldmeier-Krause17} and Mira-type long period variable star (\cite{Blum03, Dong17}, \textit{ID 552}). \textbf{SiO 19} is PSD J174541.27-290049.9 \cite{Peeples07}. It has been identified as an AGB star \cite{Blum03} with a long yearly timescale variability \cite{Dong17}, \textit{ID 299}.

We identify \textbf{SiO 24} as PSD J174541.35-290033.0 \cite{Peeples07}, which is classified as a late-type star \cite{Feldmeier-Krause17} with long timescale variability \cite{Dong17}, \textit{ID 778}. \textbf{SiO 26} is identified with PSD J174541.75-290004.6 \cite{Peeples07}. \cite{Blum03} classified it as an AGB star. \cite{Matsunaga09} and \cite{Dong17} (\textit{ID 448}) consider it a LPV star. No counterpart was found for the star \textbf{SiO 27} in the literature, and represents a new detection.

Certain similarities can be observed in the detected sources in our datasets. These sources are predominantly late-type red supergiant and AGB stars, and have been observed at different wavelengths in millimetre line emission and IR. All sources detected in the ATCA dataset are classified as M-type variable stars and have negative LSR velocities. Most of the sources exhibit long period variability with a period of few months to few years. We do not detect any YSO candidates.
\vspace{-0.2cm}
\subsection{Variability of SiO masers}

The SiO masers are associated with long period variable stars, such as AGB stars. They have been known to have variability of the timescale from few days to thousand days or more. The period of variability may not be strongly defined, and some times may be completely irregular. It is known that the SiO variability and the IR variability of a maser are correlated. The SiO and IR variabilities are related to the variation in the local heating rate due to underlying stellar pulsation, usually dominated by the radiative processes, where the IR stellar continuum radiation is absorbed by the SiO molecules in the circumstellar envelope \cite{Elitzur92, Pardo04}. Our continuous monitoring of the GC allows us to study the variability of the SiO masers in the central parsec.

\begin{figure}
  \centering
  \includegraphics{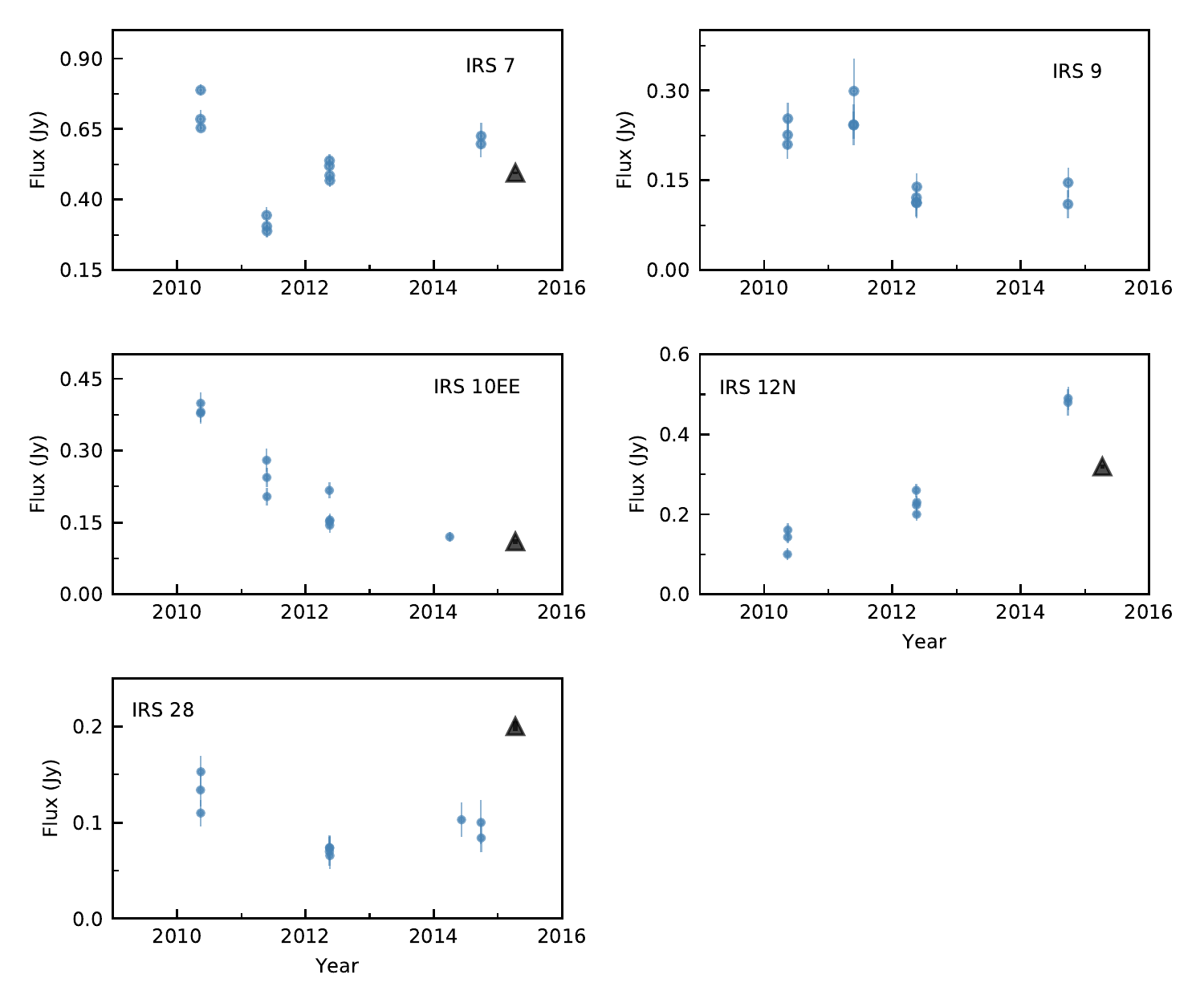}
  \caption{\small Observed variability in the peak flux of the SiO maser sources. The small blue circles represent the peak flux density observed in the ATCA dataset, while the peak flux density from the ALMA dataset is shown with black triangle. The error bars are 1$\sigma$ rms noise in the channel in which the peak flux density is observed.}\label{SiO:var}
\end{figure}

The detailed values of the flux densities observed - for each observation day when the source was detected - are given in table \ref{SiO_data:ATCA}. The flux values are consistent within the measurement accuracy and rms noise for the same observation time ($\sim$ few days), but the values can be seen to vary significantly over longer periods. Some of the sources show a periodic behaviour while others show monotonic increase or decrease in the peak flux density. Sources such as IRS 7, IRS 9, and IRS 28 exhibit a variation timescale of the order of two years, IRS 10EE shows a gradual decrease in the peak flux density. The variation in the flux density of IRS 12N suggests a possibly longer period variability. Some of these sources are classified as Mira variable stars. Mira variable stars show strong variability of the order of few hundred days where their flux density in IR may change by more than a magnitude. Since the SiO variability is directly correlated with the IR variability, strong changes in the peak flux density are also expected at the SiO transition lines. Fig. \ref{SiO:var} shows the flux density variation of the SiO maser sources seen in the ATCA and ALMA datasets over the period of the observations. Our conclusions on the variability of the sources are limited by our observation time of 5 years with an interval of $\sim 1$ year. The weaker sources that exhibit a stable flux may have a weakly defined period on a different, possibly longer timescales than our observations, or their variability may be irregular.

\vspace{-0.2cm}
\section[]{Summary}

We present our observations of the central parsec of the Galactic Centre with ATCA at 86 GHz, taken between 2010 and 2014. This is the largest dataset of GC observations at 3 mm wavelength. We detected 5 SiO maser sources, 4 have been reported at 3 mm for the first time viz. IRS 7, IRS 9, IRS 12N, and IRS 28. We also report the results from ALMA observation at 86 GHz taken on 10 April 2015. From this dataset, 20 sources were detected, of which 11 have been reported at 3 mm for the first time and 4 newly detected sources. We calculated the proper motion of these sources and the proper motion velocities are consistent with the previous observations within uncertainties. Our calculations of the proper motion though limited by resolution and sensitivity, are crucial for future high resolution and sensitivity observations and long term monitoring to constrain the proper motions of these sources and to get an insight into their orbits. We also studied the variability of the individual SiO maser sources. We detect significant variation in the flux density of the stars, though not all sources show a periodicity. We observe long period variability in the flux density of IRS 7, IRS 12N and IRS 28. Other sources show gradual increase or decrease, or irregular or weakly defined variation in the flux density.

High resolution and high sensitivity observations at 3 mm wavelength, such as with the ALMA telescope and 3 mm-VLBI observations, would be crucial in the discovery of new and weaker sources in the 30 arcsec of Sgr A*, as well as in obtaining accurate proper motion of the stellar sources in the central parsec. Further observations in different frequency bands are necessary to obtain a census of the stars in central cluster, and to discover new AGB stars, red supergiants and YSOs in that region, which will help in improving our understanding of the nature of the central stellar disk.
\vspace{-0.4cm}
\section*{Acknowledgements}

\small{This work was supported in part by the Deutsche Forschungsgemeinschaft (DFG) via the Cologne Bonn Graduate School (BCGS), the Max Planck Society through the International Max Planck Research School (IMPRS) for Astronomy and Astrophysics, as well as special funds through the University of Cologne and SFB 956 - Conditions and Impact of Star Formation, and the EU-ARC.CZ Large Research Infrastructure grant LM2015067. Part of this work was supported by fruitful discussions with members of the European Union funded COST Action MP0905: Black Holes in a Violent Universe and the Czech Science Foundation DFG collaboration (No. 13-00070J) and with members of the European Union Seventh Framework Program (FP7/2007-2013) under grant agreement no 312789; Strong gravity: Probing Strong Gravity by Black Holes Across the Range of Masses.

This paper is based on the following ALMA data: ADS/JAO.ALMA\#2013.1.00834.S. ALMA is a partnership of ESO (representing its member states), NSF (USA) and NINS (Japan), together with NRC (Canada) and NSC and ASIAA (Taiwan) and KASI (Republic of Korea), in cooperation with the Republic of Chile. The Joint ALMA Observatory is operated by ESO, AUI/NRAO, and NAOJ.}
\vspace{-0.4cm}
\bibliographystyle{JHEP}
\bibliography{paper2}

\vspace{-0.3cm}


\end{document}